\documentclass[conference,a4paper,10pt]{IEEEtran}
\usepackage{amsmath}
\usepackage{amssymb}
\usepackage{xcolor}
\usepackage{amsthm}
\usepackage{mathtools}
\usepackage{siunitx}
\usepackage{cuted}
\usepackage{graphicx}
\usepackage{subcaption}
\usepackage{epstopdf}
\usepackage{tabularx}
\usepackage{multirow}
\usepackage{booktabs}
\usepackage{diagbox}
\usepackage{cuted}
\usepackage{lipsum}

\usepackage[english]{babel}
\usepackage[utf8]{inputenc}
\usepackage{algorithm}
\usepackage{algpseudocode}

\makeatletter
\newcommand{\removelatexerror} {\let\@latex@error\@gobble}
\makeatother

\newif\iftrackrivision
\iftrackrivision
	
\else
	
\fi

\IEEEoverridecommandlockouts
\begin{document}

\title{AI-Empowered VNF Migration as a Cost-Loss-Effective Solution for Network Resilience}

\author{
	\IEEEauthorblockN{Amina~Lejla~Ibrahimpa\v{s}i\'c,~Bin~Han,~and~Hans~D.~Schotten}
	\IEEEauthorblockA{Division of Wireless Communications and Radio Positioning (WiCoN) \\Department of Electrical and Computer Engineering\\Technische Universit\"at Kaiserslautern}%
	\thanks{Bin Han (binhan@eit.uni-kl.de) is the corresponding author.}
}


\maketitle

\begin{abstract}
With a wide deployment of Multi-Access Edge Computing (MEC) in the Fifth Generation (5G) mobile networks, virtual network functions (VNF) can be flexibly migrated between difference locations, and therewith significantly enhances the network resilience to counter the degradation in quality of service (QoS) due to network function outages. A balance has to be taken carefully, between the loss reduced by VNF migration and the operations cost generated thereby. To achieve this in practical scenarios with realistic user behavior, it calls for models of both cost and user mobility. This paper proposes a novel cost model and a AI-empowered approach for a rational migration of stateful VNFs, which minimizes the sum of operations cost and potential loss caused by outages, and is capable to deal with the complex realistic user mobility patterns.
\end{abstract}


\IEEEpeerreviewmaketitle

\section{Introduction}\label{sec:intro}

Fifth-generation (5G) mobile networks are expected to extend the traditional consumer-based market with business-oriented applications. To fully exploit the potential of 5G, MEC will play a key role~\cite{barbarossa2014communicating}. MEC is intended to distribute the computing tasks and mobile services from the traditional centralized cloud server onto the edge of mobile networks, which can improve the flexibility, efficiency, timeliness, and reliability of cloud services for heterogeneous use cases. As the mobile networking service itself, driven by the emerging development of software-defined networking (SDN) and network function virtualization (NFV) technologies, will also be highly cloudified in 5G and beyond, MEC can also be deployed for virtualized network functions (VNFs) to improve the networking performance~\cite{chaudhry2020improved}. A typical  technology  of this kind, known as the VNF migration~\cite{han2017resiliency,sarrigiannis2020online}, is to create temporary edge-cloud redundancy for VNFs that are detected to be in abnormal situations that may lead to failures or malfunctions, so as to enhance the network resilience.

Compared to other pro-resilience measures such as state management and rollback recovery, VNF migration is more effective in coping with network outages caused by degraded backhaul connections. The general idea is to temporarily migrate VNFs running in the central cloud to the edge clouds, and to periodically update them. However, a general creation and maintenance of redundancy commonly for all central cloud VNFs and globally in every edge cloud will lead to a huge operation expenditure (OPEX) beyond any affordable level. It was therefore proposed in ~\cite{han20185g} to selectively migrate only the critical VNFs in outage-risky edge clouds, where the estimated loss caused by potential central cloud VNF outage, as an opportunity cost, exceeds the essential operation cost for create/update the local edge cloud redundancy.

It shall be remarked that most VNFs are {\it stateful}~\cite{yi2020design}. Therefore, besides the software implementation of the VNF itself, also the status information of the VNF must be migrated to the target cloud, so as to guarantee a seamless transfer of the service. Typical states for most stateful VNFs can be classified into control state, per-flow state, and aggregate state~\cite{han2017resiliency}. In addition to them, for some specific VNFs that are related to security and privacy, such as the virtualized authentication, authorization and accounting (V-AAA) in 5G~\cite{wong2017virtualized}, profiles of the served subscribers become also an essential state information. In this case, the migration cost is also related to the number of migrated SPs, and the cost-optimal migration control must therefore take into account the context information of subscriber devices such as position, mobility and local traffic scenario. To cope with this issue, we have proposed in~\cite{han2019context} a context-aware solution of SP synchronization. Despite of its success in demonstrating the principle, the referred work has only considered a highly simplified mobility model, i.e. a modified random walk model, which fails to characterize with the mobility pattern of users in complicated real-life environments. Towards a cost-efficient ultra-resilient MEC, in this paper we propose an AI-based solution that makes it realistic to flexibly deploy cost-optimal context-aware VNF migration in various complex scenarios of real life. 

Our work consists of four main contributions: \textit{i}) We refine the cost and loss models of VNF migration in existing work, taking into account both the user-independent part of VNF redundancy creation and the user-dependent part of SP synchronization. \textit{ii)} Based on the refined models, we derive a novel cost-loss-optimal policy of VNF migration, and an algorithm to solve it. \textit{iii)} We propose to use a mixture density neural network (MDN) to accurately model the complex user mobility pattern in reality, so as to support the prediction of subscribers' edge cloud visit probability, which is crucial in making the optimal migration decision. The proposed MDN is trained and tested with real-life GPS trajectory data of mobile subscribers. At last, \textit{iv} We carry out numerical simulations to demonstrate the effectiveness and cost efficiency of our proposed approach with realistic user mobility behaviors.
   
The remainder of this paper is organized as follows:
Sec.~\ref{sec:migration} sets up the system model, from which the cost and loss in stateful VNF migration are derived in a statistical manner. Based on the models, we propose in Sec.~\ref{sec:opt_policy} a novel cost-loss-optimal migration policy, and in Sec.~\ref{sec:ai_prediction} a MDN-based subscriber mobility prediction scheme. We then demonstrate our proposed approaches in Sec.~\ref{sec:experiments} by carrying out simulations with the real-world mobile user trajectory data. To the end, in Sec.~\ref{sec:conclusion} we close our paper with conclusions to this study and some insights about future work.

% \section{Related Work}\label{sec:related}

% Edge computing allows data created by IoT devices to be processed closer to where it is produced instead of sending it to data centers/clouds. Doing the computing closer to the edge of the network causes that the analysis of the important data happens in near real-time. Service providers may also benefit from MEC by collecting more content information from customers, such as location, interests and habits, in order to introduce new services or commercialize the data. Thus, MEC represents a key technology in enabling the 5G network evolution, since it enables IT and cloud computing capabilities at the RAN edge in a close proximity to end users and it satisfies the demanding requirements of throughout, low latency, high bandwidth, scalability and reliability.

% The process of central to edge cloud VNF migration requires periodical update of the central cloud VNF redundancies in edge clouds. If the network continuously updates all redundancies of every VNF in every edge cloud, it will lead to a huge sum of the migration cost which is economically unaffordable or at least inefficient.
% Instead, a selective redundancy update solution called the 5G Island (5GI) is applied, in which the update process is triggered only when a significant outage risk of certain VNF in certain edge cloud is detected\cite{5GArchitecture}. This project's final results on the resilience  and security concepts and developments are carried out in the framework of the 5G-MoNArch  project with direct application to the Hamburg Smart Sea Port use case.

\section{MEC VNF Migration for 5G Resilience}\label{sec:migration}

\begin{figure*}[!htbp]
	\centering
	\includegraphics[width=.7\linewidth]{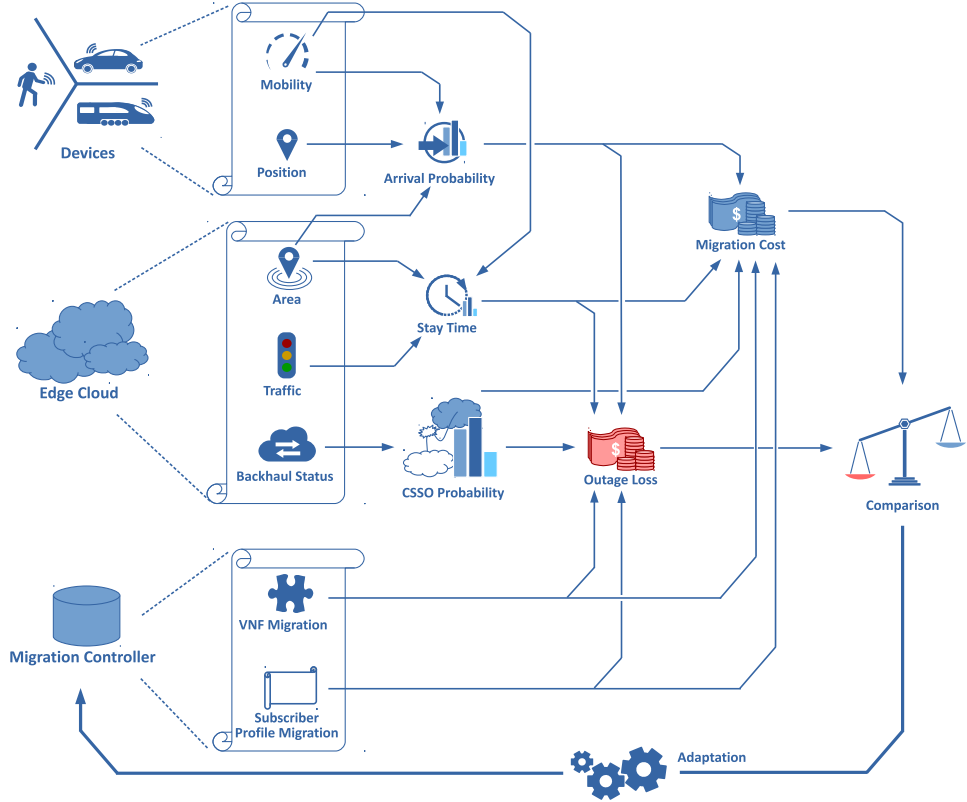}
	\caption{Concept of context-aware stateful VNF migration}
	\label{fig:concept}
\end{figure*}

\subsection{Overview}
5G networks are highly virtualized and cloudified, making it possible to flexibly deploy the software-defined networking services at variant positions in the network architecture. Especially, it has been a technical trend to distribute network functions, especially the latency-sensitive ones, to MEC servers at the edge of network. This does not only reduces the E2E latency, but also improves the reliability by relieving from the risks of congestion and malfunctions at cell site gateways and backhaul connections. Nevertheless, there are still some network functions that shall be generally located in the central cloud, including computationally challenging ones such like network-wide AI model training, and security intolerant ones such like authentication, authorization and accounting.

Concerning potential network outages that can significantly degrade the quality of service (QoS) of such centralized VNFs, it is proposed to proactively create temporary at MEC servers local redundancies for the the involved VNFs and subscriber profiles, upon the VNF outage risk. In order to estimate the outage risk - of a certain VNF, for a specific subscriber, and in the coverage of a certain edge cloud (EC) - factors that should be taken into the account are the backhaul reliability and the subscriber service time. While the backhaul reliability can be predicted by network monitoring and analyzing techniques straightforwardly, the estimation of subscriber service time relies on advanced brokering among various context information such as user motion pattern, local mobility model, EC coverage, service data traffic pattern, etc.~\cite{han2019context}.

It costs not only data traffic to update the redundancies of central cloud VNFs in edge clouds, but also server resource to maintain them. Hence, there is a trade-off between the network resilience and the operations expenditure when deploying the MEC VNF migration. Aiming at an optimum, a reasonable policy is only to migrate a central cloud VNF to MEC server, when the expected potential loss upon its outage exceeds the cost to migrate it~\cite{han20185g}.

\subsection{Model Setup of Stateful VNF Migration}
We consider a certain area - notated as the \emph{edge coverage} (EC) - that can be served by its dedicated MEC server, and a stateful VNF that by default runs in the central cloud. The MEC server keeps monitoring the QoS of the VNF, and is able to, once after every synchronization interval $T$, synchronize the VNF together with involved SPs from the central cloud to a local redundancy. Redundancies of both the VNF and the SPs will be erased at the end of the synchronization interval, in order to save maintenance cost and ensure security.  It generates a cost of $c_\text{NF}$ for every migration of the VNF, and a cost of $c_\text{SP}$ for every synchronization of every SP.

We consider the reliability $r_\text{NF}$ of the VNF in every synchronization interval $T$ to be a discrete-time random process, which can be characterized by a Markov chain~\cite{han2019context}. When it drops below a threshold $r_\text{min}$, an outage of the VNF is considered to be present in the EC. In this case, every subscriber in the EC will suffer from a degraded QoS, causing the operator a loss of $l$ per involved subscriber per unit time, unless 1) the VNF has been migrated to the MEC server, \emph{and} 2) the subscriber's profile has been synchronized to the MEC server.

Thus, at the end of an arbitrary synchronization interval, call the current time slot $t$, consider the set $\mathcal{U}(t)$ of all users in the network that may visit the target EC over the next synchronization interval $[t+1,t+T]$, the expected loss caused by the potential VNF outage is

\begin{equation}\label{eq:loss_model}
    L(t)=l\sum\limits_{\tau=t+1}^{t+T}\sum\limits_{u\in\mathcal{U}(t)}p_\text{o}(\tau)p_{\text{v},u}(\tau)\left[1-m(t)s_u(t)\right],
\end{equation}
where $p_o(\tau)=\text{Prob}\left[r(\tau)<r_\text{min}\right]$ is the risk of VNF outage in time slot $\tau$, $p_{\text{v},u}(\tau)$ is the probability that user $u$ visits EC in time slot $\tau$. $m(t)$ is the VNF migration label that has the value of $1$ when the VNF redundancy is created/updated in interval $[t+1,t+T]$, and equals $0$ otherwise. Similarly, $s_u(t)$ is the SP synchronization label for user $u$. 
Correspondingly, the total cost generated by the migration decision is
\begin{equation}\label{eq:cost_model}
    C(t)= c_\text{NF}m(t)+c_\text{SP}\sum\limits_{u\in\mathcal{U}(t)}s_u(t).
\end{equation}

To achieve a rational decision of VNF migration with SP synchronization, the migration controller must predict the key probabilities $p_\text{o}(\tau)$ and $p_{\text{v},u}(\tau)$ for all $u\in\mathcal{U}$ and $\tau\in[t+1,t+T]$ upon the context information provisioned by VNF monitoring and mobility management, as illustrated in Fig.~\ref{fig:concept}.

\section{Cost-Loss-Optimal Stateful VNF Migration}\label{sec:opt_policy}
For convenience of notation let $T_u(t)\triangleq\sum\limits_{\tau=t+1}^{t+T}p_\text{o}(\tau)p_{\text{v},u}(\tau)$, which is the VNF outage time that user $u$ is expected to experience over the next synchronization interval $[t+1,t+T]$. Combining the models of outage loss~\eqref{eq:loss_model} and migration cost~\eqref{eq:cost_model}, it can be obtained for any certain VNF migration decision $m(t)$ and any certain SP synchronization plan $\mathcal{U}_s(t)\triangleq\{u: s_u(t)=1, \forall u\in\mathcal{U}(t)\}$, that the expected sum of cost and loss over the next interval is

\begin{strip}
%\noindent\rule{\textwidth}{0.4pt}
\begin{equation}\label{eq:cost_loss_sum}
\begin{split}
    S(t)=&L(t)+C(t)=c_\text{NF}m(t)+\sum\limits_{u\in\mathcal{U}(t)}\left\{l\sum\limits_{\tau=t+1}^{t+T}p_\text{o}(\tau)p_{\text{v},u}(\tau)\left[1-m(t]s_u(t)\right)+c_\text{SP}s_u(t)\right\}\\
    =&\begin{cases}
    c_\text{NF}+\sum\limits_{u\in\mathcal{U}(t)}\left\{lT_u(t)+s_u(t)\left[c_\text{SP}-lT_u(t)\right]\right\}&\text{if}~m(t)=1\\
    \sum\limits_{u\in\mathcal{U}(t)}\left[T_u(t)+c_\text{SP}s_u(t)\right]&\text{if}~m(t)=0
    \end{cases}.
\end{split}
\end{equation}
%\noindent\rule{\textwidth}{0.4pt}
\end{strip}

Especially, in the case of $m(t)=1$, i.e. the VNF is migrated to the MEC server, we have the lower bound
\begin{equation}\label{eq:migration_lower_bound}
    S\left(t\vert m(t)=1\right)\geqslant c_\text{NF}+\sum\limits_{\substack{u\in\mathcal{U}(t)\\lT_u(t)<c_\text{SP}}}lT_u(t)+\sum\limits_{\substack{u\in\mathcal{U}(t)\\lT_u(t)\geqslant c_\text{SP}}}c_\text{SP},    
\end{equation}
where the equality holds when and only when
\begin{equation}\label{eq:opt_sync}
    s_u(t)=s_u^\text{opt}\left(t~\vert~m(t)=1\right)=\begin{cases}
            1&lT_u(t)\geqslant c_\text{SP}\\
            0&\text{otherwise}
        \end{cases},
\end{equation}
which suggests to synchronize the UPs only for the users that are risking outage losses higher than the UP migration cost.

On the other hand, when $m(t)=0$, i.e. the VNF is not migrated, we have another lower bound
\begin{equation}\label{eq:no_migration_lower_bound}
    S\left(t~\vert~m(t)=0\right)\geqslant\sum\limits_{u\in\mathcal{U}(t)}lT_u(t),
\end{equation}
where the equality holds when and only when
\begin{equation}\label{eq:no_sync}
    s_u(t)=s_u^\text{opt}\left(t~\vert~m(t)=0\right)=0,\forall u\in\mathcal{U}(t),
\end{equation}
suggesting to synchronize no UP, which is trivial since the UPs are only made use of when the VNF is migrated.

Obviously, a cost-loss-optimal migration policy can be achieved by comparing the two achievable bounds \eqref{eq:migration_lower_bound} and \eqref{eq:no_migration_lower_bound}, choosing the VNF migration decision that leads to the lower one, and the corresponding SP synchronization plan by \eqref{eq:opt_sync} and \eqref{eq:no_sync}, respectively. By means of this, we propose the cost-loss-optimal stateful VNF migration algorithm, as described in Algorithm~\ref{alg:cost_opt_migration}.

\begin{algorithm}[!htbp]
\caption{Cost-Loss-Optimal Stateful VNF Migration}
\label{alg:cost_opt_migration}
\begin{algorithmic}[1]
\State \textbf{Input} $\mathcal{U}(t), l, c_\text{NF}, c_\text{SP}, T, r_\text{NF}(t)$
\item[]
\State $S_1\gets c_\text{NF}$\Comment{To buffer the lower bound \eqref{eq:migration_lower_bound}}
\State $S_2\gets 0$\Comment{To buffer the lower bound \eqref{eq:no_migration_lower_bound}}
\State $\mathcal{U}_s(t)\gets\emptyset$
\For {$u \in \mathcal{U}(t)$}
    \State $T_u(t)\gets 0$
\EndFor
\item[]
\For{$\tau \in [t+1, t+T]$}
    \State  $p_\text{o}(\tau) \gets \texttt{EstVNFOutageProb}(r_\text{NF}(t),\tau-t)$
    %\State \Comment{\parbox[t]{.83\linewidth}{\emph{The VNF outage probability can be estimated according to the real-time monitored status and Markov model of VNF reliability.}}}
    %\item[]
    \For {$u \in \mathcal{U}(t)$}
        \State $p_{\text{v},u}(\tau) \gets \texttt{PredECVisitProb}(u, \tau-t)$
        %\State \Comment{\parbox[t]{.83\linewidth}{\emph{The prediction of users' EC visiting probabilities relies on context awareness}}}
        %\item[]
        \State $T_u(t)\gets T_u(t)+p_\text{o}(\tau)p_{\text{v},u}(\tau)$
    \EndFor
\EndFor
\item[]
\For {$u \in \mathcal{U}(t)$}
    \State $S_2\gets S_2+lT_u(t)$
    \If{$lT_u(t)\geqslant c_\text{SP}$}
    \State $\mathcal{U}_s(t)\gets\mathcal{U}_s(t)\cup\{u\}$
    \State $S_1\gets S_1+c_\text{SP}$
    \Else
    \State $s_u(t)\gets 0$
    \State $S_1\gets S_1+lT_u(t)$
    \EndIf
\EndFor
\item[]
\If{$S_1\geqslant S_2$}
    \State $m(t)\gets 1$
\Else
    \State $m(t)\gets 0$
    \State $\mathcal{U}_s(t)\gets\emptyset$
\EndIf
\State \textbf{Return} $m(t),\mathcal{U}_s(t)$
\end{algorithmic}
\end{algorithm}

\section{MDN to Model and Predict User Mobility}\label{sec:ai_prediction}

\begin{figure}[!htbp]
	\centering
	\includegraphics[width=\linewidth]{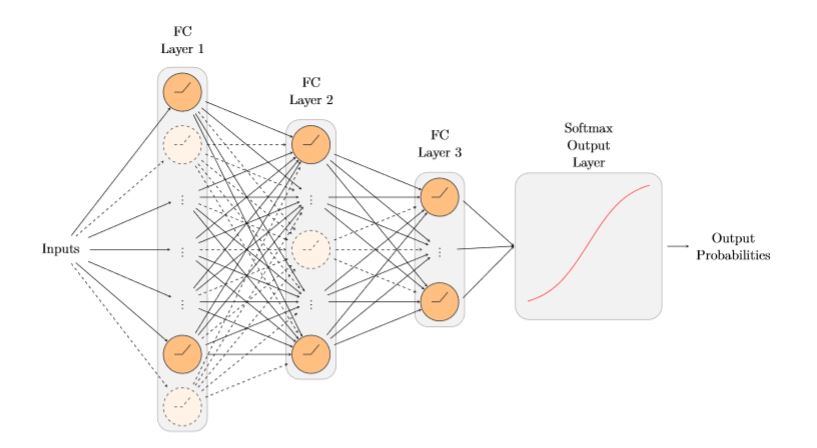}
	\caption{The structure of proposed MDN}
	\label{fig:ann}
\end{figure}

The estimation of VNF outage probability $p_\text{o}(\tau)$, as referred by the function $\texttt{EstVNFOutageProb}(~)$ in Algorithm~\ref{alg:cost_opt_migration}, can be realized by online VNF status analysis~\cite{han2019context}. Nevertheless, the prediction of subscribers' EC visit probability $p_{\text{v},u}(\tau)$, which is referred in Algorithm~\ref{alg:cost_opt_migration} by the function $\texttt{PredECVisitProb}(~)$, remains an open issue. Mainstream mobility management solutions commonly predict only  the most likely future trace, while a spatial probability distribution of the subscriber's future position is required for an accurate multi-step $p_{\text{v},u}$ prediction. To address this challenge, we invoke the mixture density network~\cite{bishop1994mixture}.

\subsection{Mixture Density Network}
The core principle of MDN is to use the outputs of a neural network to parametrize a mixture distribution. A subset of the outputs are used to define the mixture weights, while the remaining outputs are used to parametrize the individual mixture components~\cite{graves2013generating}. The probability density of the target data is then represented as a linear combination of kernel functions in the form
\begin{equation}\label{eq:mdn}
	p\left(\mathbf{t}\vert\mathbf{x}\right)=\sum\limits_{i=1}^I\alpha_i(\mathbf{x})\theta_i\left(\mathbf{t}\vert\mathbf{x}\right),
\end{equation}
where $I$ is the number of components (kernels) in the distribution mixture, the mixing coefficient $\alpha_i(\mathbf{x})$ represent the prior probability of the target vector $\mathbf{t}$ conditioned on the input status $\mathbf{x}$ generated from the $i^\text{th}$ kernel.

For any given value of x, the mixture model \eqref{eq:mdn} provides a general model for an arbitrary conditional density function $p\left(\mathbf{t}\vert\mathbf{x}\right)$. Considering the mixing coefficients $\alpha_i(\mathbf{x})$, means $\mu_i(\mathbf{x})$ and variances $\sigma_i(\mathbf{x})$ of all kernels $i\in\{1,2,\dots, I\}$, to be general (continuous) functions of $\mathbf{x}$, a conventional feed-forward neural network can be constructed, taking $\mathbf{x}$ as input and generates $p\left(\mathbf{t}\vert\mathbf{x}\right)$ as output, which is known as a mixture density network.

By choosing a mixture model with a sufficient number of kernel functions, and a neural network with a sufficient number of hidden units, a MDN can be trained to approximate as closely as desired any conditional probability distributions. In this work we propose to compose a network with 3 successive fully connected hidden layers with a rectified linear unit (ReLU) activation functions, whose output will be fed into a final softmax output layer for sampling from a mixture of bivariate Gaussians,  as illustrated in Fig.~\ref{fig:ann}.

\subsection{Real-life Mobility Modeling with MDN}\label{subsec:mdn_training}
\subsubsection{Data Provisioning}
To train the MDN towards realistic user mobility pattern, we obtained real-life user mobility data from a GPS trajectory dataset, which was collected by the \emph{Microsoft Research Asia} as a group of \emph{Geolife} project. It contains the data of 182 users in a period over five years. Every GPS trajectory in this dataset is represented by a sequence of time-stamped 3D coordinates, each containing the latitude, longitude and altitude. These trajectories were recorded by different GPS loggers and GPS phones, and have a variety of sampling rates~\cite{geolife}. 

\subsubsection{Data Pre-processing}
First of all, since we focus on the scenario of terrestrial networks, the 3D $X-Y-Z$ coordinates were first projected into the 2D $X-Y$ plane and reorganized to unified sampling rate. Afterwards, as users in various mobility classes can have significantly different behaviors, we selectively picked only the pedestrian-like trajectories by filtering high-speed users out. Furthermore, user mobility is usually time-varying, and therefore exhibits low stationary over long terms, which can critically reduce the performance of most time series analyzing techniques, including the MDN training (which is the mobility model fitting) in our case. To cope with this issue, long-term trajectories were segmented into short ones, so as to guarantee a stationarity over $92\%$ in both $X$ and $Y$ domains. 

\subsubsection{Specification and Training of the MDN}
We specified our MDN to 64 input nodes, taking the 1-order differentiated 2D coordinates over a time window of 32 samples, and 2 output nodes for a single-step forward prediction. The three fully connected hidden layers are specified with 512, 128, and 12 neurons, respectively.

We split the dataset into two parts, one with $90\%$ trajectories as the training set, and the other with the remaining $10\%$ data as the validation set. We trained the model for 15 epochs, each referring to a full traversal through the training dataset with a batch of 512 trajectories in parallel. We chose \texttt{RMSprop} as the optimizer, with the mean square error (MES) selected as loss function for fitting performance evaluation.

\begin{figure}
	\centering
	\begin{subfigure}{\linewidth}
		\centering
		\includegraphics[width=.8\linewidth]{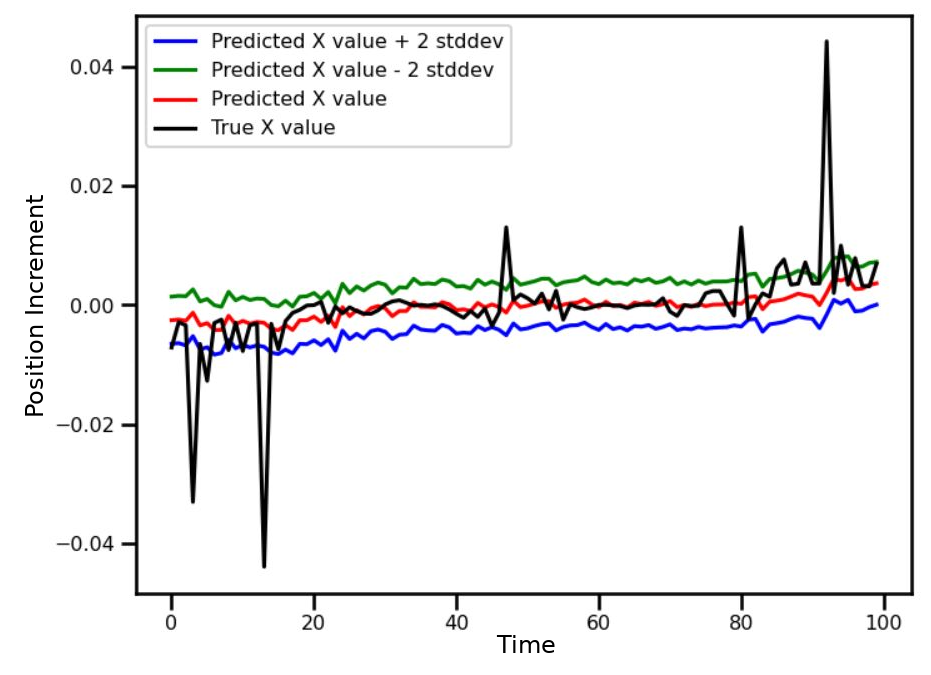}
		\caption{A sample prediction by the trained MDN}
		\label{fig:sample_prediction}
	\end{subfigure}
	\begin{subfigure}{\linewidth}
		\centering
		\includegraphics[width=.8\linewidth]{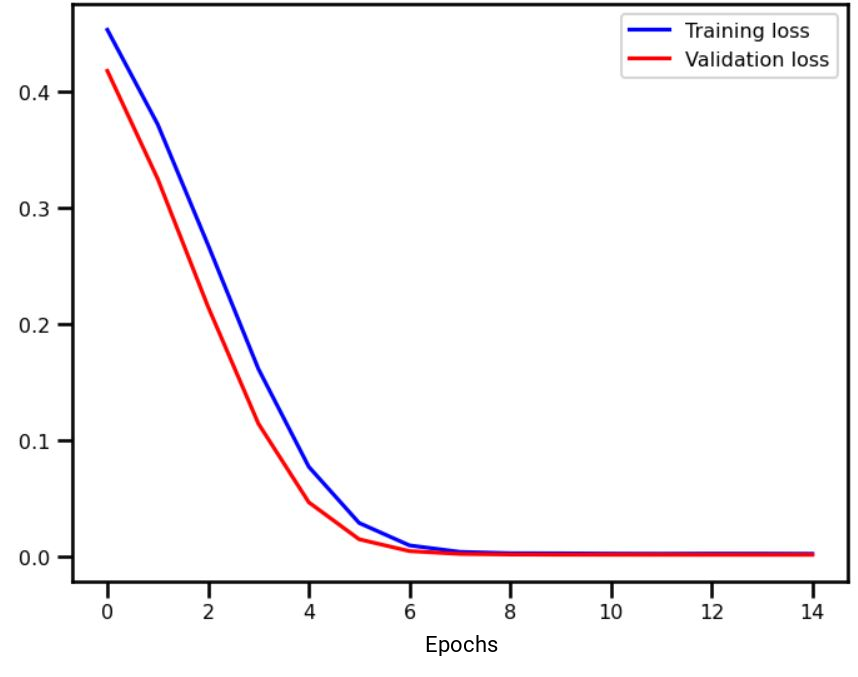}
		\caption{Training and validation losses}
		\label{fig:mdn_losses}
	\end{subfigure}
\caption{Performance of MDN-based user mobility modeling}
\end{figure}

A 100-step segment of prediction result is shown in Fig.~\ref{fig:sample_prediction} as an validation example, showing a sufficient accuracy. Both the training and validation losses drop rapidly along with the epochs, as shown in Fig.~\ref{fig:mdn_losses}, exhibiting a satisfactory convergence.

\section{Numerical Evaluation}\label{sec:experiments}
\subsection{Simulation Setup}
To evaluate the effectiveness of our proposed approach, we carried out numerical simulations. 
 
For the environment, we considered an 8x8 $km^{2}$ region, where the circular EC area with radius of 2 km in the middle is served by a MEC server. As initialization, we randomly set 1000 users across the whole region under a uniform distribution, each user was randomly assigned to a fitted kernel we had obtained during the MDN training in Sec.~\ref{subsec:mdn_training}, as its mobility model. Then for a pre-convergence from the random initial state to a steady state, we let all users randomly move w.r.t. their models for $250$ steps, with each step interval lasting \SI{1}{\minute}. If any user leaves the region, a new arriving user will be randomly created at the region boundary, so that the user density remains consistent. Meanwhile, the VNF outage events was simulated by a Markov model inherited from \cite{han2019context}, which takes account of multiple factors in VNF reliability, such as the random network status fluctuation, the rare disaster events, and the repairing procedure. The migration interval was set to $30$ user motion steps.
 
For the VNF migration controlling system, we considered a MDN to be created at then end of user location pre-convergence, and trained for the next $500$ steps to learn the user mobility pattern. With the MDN sufficiently trained, the simulation continued for another period of 4000 steps, during which the MDN continuously predicted the EC visit probability $p_{\text{v},u}$ for every user $u$ and kept online updated. Meanwhile, a VNF monitoring engine was considered to online observe the VNF status, and predict the outage risk $p_\text{o}$ at every step with a full knowledge about the outage model.

\subsection{Benchmarking}
\begin{figure}[!hbtp]
	\centering
	\includegraphics[width=\linewidth]{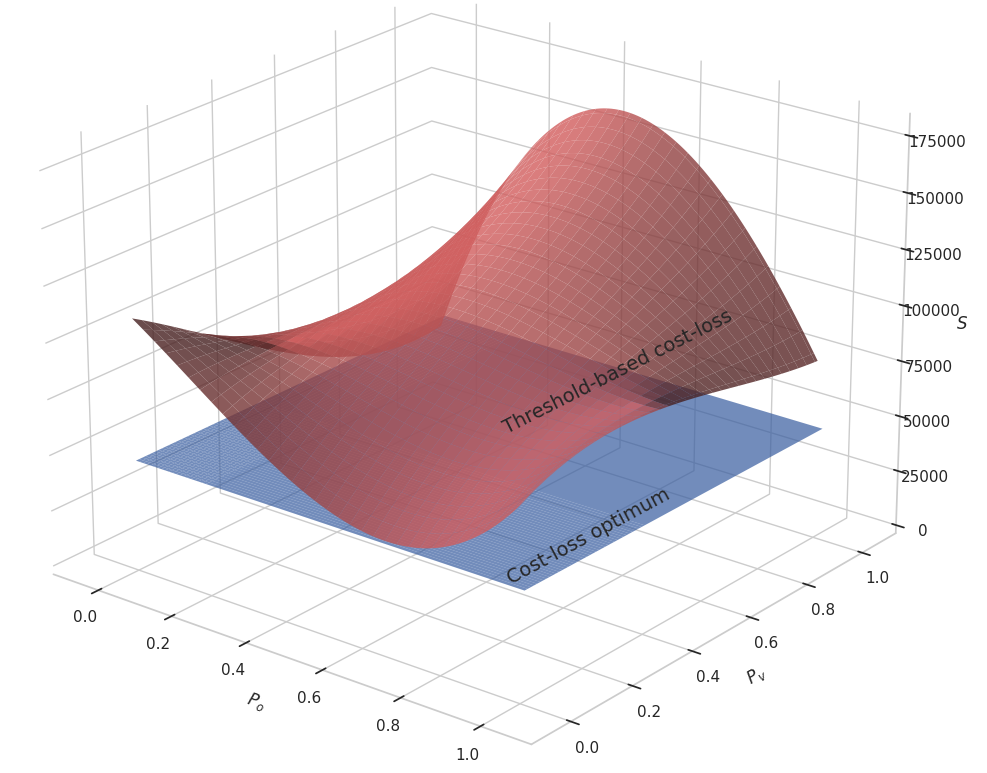}
	\caption{Result of the benchmarking}
	\label{fig:benchmark}
\end{figure}

For benchmarking we adopted the risk thresholding approach used in \cite{han2019context} with extension. Differing from \cite{han2019context} that only considers the UP synchronization cost $c_\text{UP}$, in this work we also need to take account of the VNF migration cost$c_\text{NF}$, so a double thresholding method was designed, where the controller chooses to migrate the VNF in EC only when the average $p_\text{o}$ in next synchronization interval exceeds a threshold $P_\text{o}$, and synchronizes the SP of a user only when its cumulative probability of visiting EC in the next interval exceeds another threshold $P_\text{v}$. We tested this basedline solution with different specifications of $[P_\text{o}, P_\text{v}]$, and compare the resulted sum of loss and cost to the performance achievable by the MDN-based optimal migration scheme. The results, as shown in Fig.~\ref{fig:benchmark}, shows that our proposed method outperforms the benchamrk, even when the latter is optimally specified.

\section{Conclusion}\label{sec:conclusion}
In this work, we have developed a novel cost-loss model of stateful VNF migration, concerning costs of both the VNF migration cost and the subscriber profile synchronization, and therewith proposed a cost-loss-optimal migration strategy. To cope with the challenge of predicting subscribers' edge cloud visit probability with realistic user mobility pattern, we have developed an artificial intelligence solution based on mixture density network. The proposed approach has been validated by numerical simulations.

\ifCLASSOPTIONcaptionsoff
  \newpage
\fi

\bibliographystyle{IEEEtran}
\bibliography{library}

%\begin{thebibliography}{}
%\bibitem{HNHS2019comprehensive}
%M.~A.~Habibi, M.~Nasimi, B.~Han and H.~D.~Schotten, ``A comprehensive survey of RAN architectures toward 5G mobile communication system,'' in \emph{IEEE Access}, vol. 7, pp. 70371--70421, 2019, doi: 10.1109/ACCESS.2019.2919657.
%
%\bibitem{HWM+2017security}
%B.~Han, S.~Wong, C.~Mannweiler, M.~Dohler and H.~D.~Schotten, ``Security trust zone in 5G networks,'' in \emph{2017 24th International Conference on Telecommunications (ICT)}, Limassol, 2017, pp. 1--5, doi: 10.1109/ICT.2017.7998270.
%
%\bibitem{HCS2018security}
%B.~Han, M.~R.~Crippa, and H.~D.~Schotten, ``5G island for network resilience and autonomous failsafe operations,'' in \emph{2018 European Conference on Networks and Communications (EuCNC), Special Session on 5G Mobile Network Architecture and New Radio Advances (5GMoNANeRA)}, available at 	arXiv:1805.01715 [cs.NI]
%
%\bibitem{KS2018ammcoa}
%J.~Kochems and H.~D.~Schotten, ``AMMCOA - Nomadic 5G private networks,'' in \emph{Proc. 23rd VDE/ITG Fachtagung Mobilkommunikation}, Osnabr\"uck, Germany, May 2018, pp. 105-–108.
%
%\bibitem{HWM+2019context}
%B.~Han, S.~Wong, C.~Mannweiler, M.~R.~Crippa and H.~D.~Schotten, ``Context-awareness enhances 5G multi-access edge computing reliability,'' in \emph{IEEE Access}, vol. 7, pp. 21290--21299, 2019, doi: 10.1109/ACCESS.2019.2898316.
%
%
%\bibitem{anagnostopoulos2009predicting}
%Anagnostopoulos, Theodoros, et al. "Predicting the location of mobile users: a machine learning approach." Proceedings of the 2009 international conference on Pervasive services. 2009.
%\end{thebibliography}

\end{document}